\documentclass[conference]{IEEEtran}
\IEEEoverridecommandlockouts

\usepackage{cite}
\usepackage{amsmath,amssymb,amsfonts}
\usepackage[linesnumbered,ruled,vlined]{algorithm2e}
\usepackage{graphicx}
\usepackage{textcomp}
\usepackage{xcolor}
\usepackage{amsmath}
\usepackage{multirow}
\usepackage{booktabs}
\usepackage[caption=false,font=footnotesize]{subfig}

\usepackage[hidelinks]{hyperref}
\usepackage{orcidlink} 

\def\BibTeX{{\rm B\kern-.05em{\sc i\kern-.025em b}\kern-.08em
    T\kern-.1667em\lower.7ex\hbox{E}\kern-.125emX}}
\begin{document}

\title{Policy–Value Guided MDP–MCTS Framework for Cyber Kill-Chain Inference}

\author{
\IEEEauthorblockN{Chitraksh Singh\,\orcidlink{0009-0000-1020-8989}} 
\and
\IEEEauthorblockN{Monisha Dhanraj\, \orcidlink{0009-0009-8593-1632}}
\and
\IEEEauthorblockN{Ken Huang\, \orcidlink{0009-0004-6502-3673}}
}

\maketitle
\begin{abstract}
Threat analysts routinely rely on natural-language reports that describe attacker actions without enumerating the full kill chain or the dependencies between phases, making automated reconstruction of ATT\&CK-consistent intrusion paths a difficult open problem. We propose a reasoning framework that infers complete seven-phase kill chains by coupling phase-conditioned semantic priors from Transformer models with a symbolic Markov Decision Process and an AlphaZero-style Monte Carlo Tree Search guided by a Policy–Value Network. The framework enforces semantic relevance, phase cohesion, and transition plausibility through a multi-objective reward function while allowing search to explore alternative interpretations of the CTI narrative. Applied to three real intrusions FIN6, APT24, and UNC1549 the approach yields kill chains that surpass Transformer baselines in semantic fidelity and operational coherence, and frequently align with expert-selected TTPs. Our results demonstrate that combining contextual embeddings with search-based decision-making offers a practical path toward automated, interpretable kill-chain reconstruction for cyber defense.
\end{abstract}

\begin{IEEEkeywords}
Cyber Kill Chain, MITRE ATT\&CK, Threat Intelligence, Markov Decision Process, Monte Carlo Tree Search, Policy–Value Network, Adversarial Reasoning
\end{IEEEkeywords}

\section{Introduction}

Cyber threat intelligence (CTI) reports remain one of the richest sources of information about real intrusions, yet they are written for human consumption and rarely articulate the full sequence of attacker decisions. Analysts must reconstruct the underlying kill chain from scattered textual cues, implicit dependencies, and partially described actions an increasingly difficult task as modern intrusion campaigns grow in scale, complexity, and stealth. Automated systems capable of transforming such unstructured narratives into structured, ATT\&CK-aligned attacker pathways could substantially improve incident response, post-incident forensics, and threat hunting operations. However, this problem is fundamentally challenging because CTI text describes intent, context, and constraints rather than enumerating explicit phase transitions.

Prior research has investigated decision-theoretic models as a way to reason about multi-stage adversarial behavior. Markov Decision Processes (MDPs) capture how an intrusion unfolds as a sequence of state transitions governed by probabilistic dependencies, enabling systems to reason about attacker progression even when observations are noisy or incomplete. Luo et al.~\cite{LUO2025480} showed that such models can support real-time detection under dynamic conditions, while Liu et al.~\cite{LIU2021102480} extended this idea using partially observable MDPs and deep reinforcement learning to handle uncertainty in strategic attacker choices. These studies demonstrate that structured probabilistic models can generalize beyond static signatures and adapt to evolving adversarial strategies.

In parallel, Monte Carlo Tree Search (MCTS) has emerged as an effective mechanism for navigating large decision spaces when exhaustively enumerating all possible attack paths is infeasible. Work by Sartea et al.~\cite{SARTEA2020103303} highlights how simulation-driven search can uncover behavioral pathways not visible to static analysis, underscoring the value of exploration in settings where attacker behavior is branching and long-horizon. More recent efforts integrate learning-based priors with sequential search to capture richer threat dynamics. Markov games have modeled adversarial interactions across attackers and defenders~\cite{alavizadeh2021markovgamemodelaibased}, while planning under imperfect observability has been applied to deception~\cite{fu2022almostsureintentiondeceptionplanning} and stealthy decision-making~\cite{wei2025planningstealthybackdoorattacks}. Collectively, these works motivate architectures that combine semantic understanding, probabilistic modeling, and guided exploration \cite{lakshminarayana2017optimalattackcyberphysicalcontrol}.

Our prior system, KillChainGraph~\cite{singh2025killchain}, demonstrated that supervised learning can map CTI text to MITRE ATT\&CK techniques with reasonable accuracy, but its outputs were static and phase-local. It could not reason about how techniques depend on one another or construct coherent multi-step adversarial pathways. As a result, it lacked the ability to approximate the analytical reasoning performed by human threat hunters when reconstructing full kill chains.

The present work addresses this gap by introducing a unified reasoning framework that infers a complete seven-phase ATT\&CK kill chain from unstructured text. The approach integrates three components: (1) Transformer-based phase classifiers that provide phase-conditioned semantic priors over techniques; (2) a multi-objective reward model that scores partial attack paths using semantic relevance, cross-phase cohesion, empirical transition likelihoods, and defensive exposure costs; and (3) a Policy--Value Network trained on synthetic MDP trajectories to produce context-aware policies and value estimates for symbolic search. These components are combined through an AlphaZero-style MCTS procedure, enabling the system to explore plausible adversarial trajectories while grounding decisions in both narrative semantics and structural constraints.

By integrating contextual embeddings with a search procedure grounded in reinforcement-style decision dynamics, the framework moves beyond static classification and toward an interpretable, sequential reasoning process. The inferred kill chains capture cues that are expressed implicitly within the CTI narrative, exhibit alignment with characteristic adversary tradecraft, and preserve coherence across successive ATT\&CK phases. Empirical results indicate that the system frequently identifies transitions consistent with those chosen by human analysts, though gaps remain in settings where narrative signals are diffuse or operational intent is ambiguous. Overall, the findings suggest that combining semantic priors with search-guided symbolic reasoning offers a principled and extensible basis for automated reconstruction of attacker workflows.

\begin{figure}[htbp]
    \centering
    \includegraphics[width=0.95\linewidth]{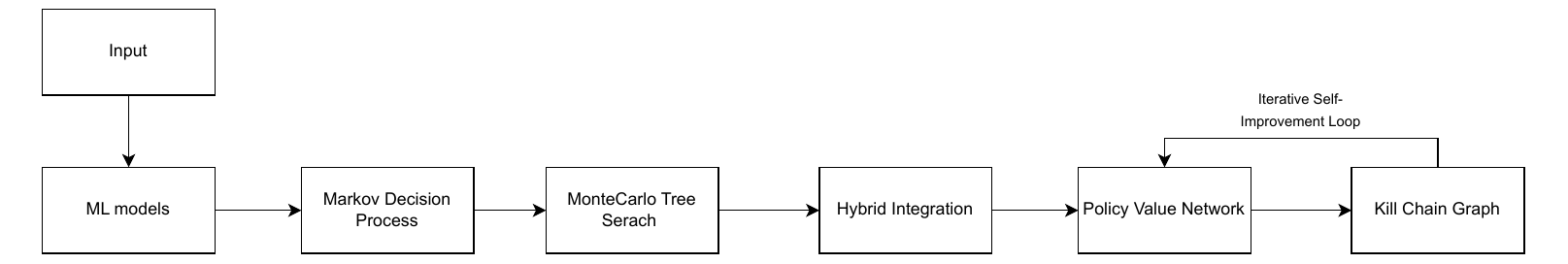}
    \caption{Proposed Cyber Kill-Chain Reasoning Framework.}
    \label{fig:framework_architecture}
\end{figure}

\section{Related Work}

Research on modeling adversarial behavior has increasingly turned toward decision theoretic, simulation driven, and learning based approaches to capture the complexity of modern intrusion campaigns. These efforts span Markov Decision Process (MDP) formulations of cyber operations, Monte Carlo Tree Search (MCTS) for exploratory reasoning, reinforcement learning for adaptive defense, and broader system-theoretic perspectives on cybersecurity as an evolving process.

MDPs have been widely adopted to represent sequential dependencies in cyberattacks and to support adaptive detection and response. Luo et al.~\cite{LUO2025480} demonstrated that MDP driven detection systems can update action policies in real time, outperforming static deep learning architectures when network conditions evolve. Liu et al.~\cite{LIU2021102480} incorporated a partially observable MDP with a recurrent deep Q-network to reason about attacker defender interactions under uncertainty, achieving robust operational performance in dynamic environments. Complementary work by Zhou et al.~\cite{MDPautomatecybersec} framed automated cyber defense as a dynamic MDP with evolving reward structures, while Li and Wu~\cite{Li2025} introduced a prior-informed Proximal Policy Optimization method that exploits domain knowledge for efficient pruning during attack-graph traversal. Collectively, these contributions underscore the viability of probabilistic sequential models for reasoning about multi-stage adversarial behavior.

MCTS has been explored as a way to navigate large or partially observable attack spaces where exhaustive enumeration is infeasible. Sartea et al.~\cite{SARTEA2020103303} applied MCTS to active Android malware analysis, showing that guided simulation can uncover behavioral pathways missed by conventional static methods. MCTS is particularly suited for intrusion analysis because it allocates computational effort toward uncertain but promising branches, allowing systems to explore attacker strategies without requiring fully specified attack graphs.

Hybrid models that combine probabilistic reasoning, game theoretic representations, and simulation driven exploration have further expanded the applicability of these methods. Alavizadeh et al.~\cite{alavizadeh2021markovgamemodelaibased} used Markov games with deep policy learning to capture adaptive interactions in attack defense settings. Fu~\cite{fu2022almostsureintentiondeceptionplanning} and Wei et al.~\cite{wei2025planningstealthybackdoorattacks} investigated deception and stealth in adversarial planning, demonstrating that MDP based reasoning augmented with search can approximate strategic, multi-step attacker decision processes under imperfect observability.

Beyond algorithmic techniques, system level perspectives highlight the dynamic and interdependent nature of cybersecurity. Xu’s Cybersecurity Dynamics framework~\cite{cybersecdynamics} characterizes security as an evolving process shaped by feedback loops and interacting system states. Collins~\cite{thinkforcybersec} introduced a systems thinking methodology emphasizing cognitive and structural reasoning to interpret complex attack ecosystems. These perspectives reinforce the need for analytical frameworks that capture temporal evolution, semantic relationships, and probabilistic dependencies.

Despite this progress, most existing approaches operate within predefined attack graphs or simulated environments and therefore do not address the challenge of constructing kill-chain narratives directly from unstructured CTI text. Systems such as KillChainGraph~\cite{singh2025killchain} demonstrated that supervised learning can map portions of CTI narratives to individual ATT\&CK techniques, but they do not extend to multi-phase reasoning or exploratory inference across full attack sequences. The framework introduced in this study advances this direction by integrating Transformer based semantic priors, an MDP formulation of phase-wise transitions, a multi-objective reward function, a Policy-Value Network for contextual policy and value estimation, and an MCTS component for exploration. Through this combination, the system produces structured kill-chain hypotheses that reflect both linguistic evidence in threat reports and probabilistic reasoning over attacker trajectories.

\begin{figure}[t]
    \centering
    \includegraphics[width=\linewidth]{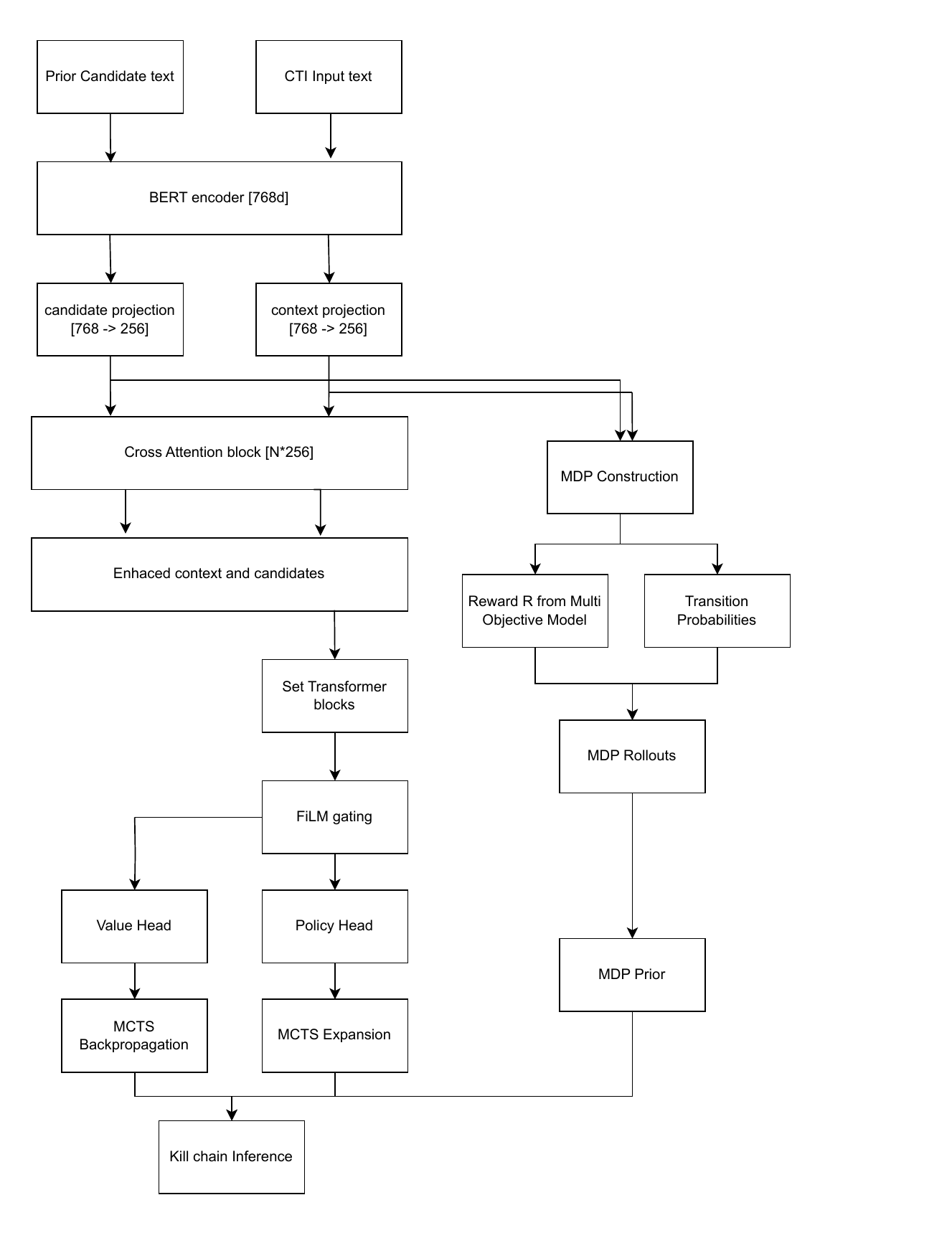}
    \caption{Reasoning pipeline combining semantic encoding, multi-objective rewards, MDP rollouts, PV-Network predictions, and MCTS inference.}
    \label{fig:pipeline_diagram}
\end{figure}

\section{Methodology}

The goal of this framework is to infer a coherent seven-phase ATT\&CK kill chain directly from an unstructured cyber threat intelligence (CTI) report as shown in Fig.\ref{fig:framework_architecture}. This requires combining semantic interpretation of the narrative, structural reasoning about plausible adversarial progressions, and search-based optimization over symbolic techniques. Our approach integrates five components: (1) contextual semantic embeddings derived from Transformer encoders, (2) a multi-objective reward measuring semantic and operational plausibility, (3) an MDP that models symbolic phase-to-phase transitions, (4) a Policy-Value Network (PVN) that estimates context-sensitive priors and values, and (5) a Monte Carlo Tree Search (MCTS) procedure that unifies learning, reward, and symbolic structure into a single inference mechanism. The overall flow is illustrated in Fig.\ref{fig:pipeline_diagram}.

\subsection{Dataset and Semantic Embedding Layer}

All symbolic reasoning is in the MITRE ATT\&CK enterprise matrix. Each technique $t$ provides a unique identifier, natural language description, associated tactic, and metadata including detection sources and mitigation guidelines. These descriptions anchor the symbolic space in natural language and permit semantic alignment with CTI reports.

A BERT encoder maps the CTI report $x$ into a contextual vector $c\in\mathbb{R}^{768}$ capturing threat intent, operational goals, and indicators referenced in the narrative. The same encoder converts each technique description into an embedding $v_t\in\mathbb{R}^{768}$. To enable joint reasoning, both $c$ and $v_t$ are projected into a shared $256$-dimensional latent space, where contextual and candidate semantics can interact.

Bidirectional cross-attention is applied to bind the CTI narrative to candidate techniques. The context-to-candidate attention highlights techniques whose semantics align with the threat description, while the candidate-to-context attention extracts global contextual cues that modulate subsequent reasoning. A Set Transformer block then models higher-order interactions among the candidate techniques. Unlike standard Transformers, Set Transformers use permutation-invariant attention layers, making them naturally suited for unordered candidate sets such as the ATT\&CK techniques available at each phase. This ensures that semantic relationships among possible actions are modeled independently of their ordering.

Finally, FiLM-style modulation applies an affine transformation to each candidate embedding, parameterized by the contextual representation. This allows the CTI narrative to warp the candidate space, emphasizing techniques that are contextually plausible and suppressing those incompatible with the described threat scenario. These enhanced embeddings serve as inputs to both the Policy--Value Network and the MDP/MCTS reasoning layers.

\subsection{Multi-Objective Reward}

To evaluate partial symbolic paths, we define a multi-objective reward function that integrates semantic relevance, phase-to-phase coherence, transition likelihood, coverage of contextual priors, stealth, and exposure-related penalties. For a partial path $p_m=(t_{\phi_1},\dots,t_{\phi_m})$, the reward is decomposed as $R(p_m)=R^{+}(p_m)-R^{-}(p_m)$.

Relevance measures alignment between the CTI context and selected techniques:
\[
\text{Rel}(p_m)=\frac{1}{m}\sum_{i=1}^{m}\cos(c,v_{t_i}).
\]

Cohesion measures semantic coherence between successive phases:
\[
\text{Coh}(p_m)=\frac{1}{m-1}\sum_{i=1}^{m-1}\cos(v_{t_i},v_{t_{i+1}}).
\]

Transition plausibility reflects structural coherence induced by the MDP transition kernel:
\[
\text{Trans}(p_m)=\prod_{i=1}^{m-1}P_{\text{MDP}}(t_{i+1}\mid t_i).
\]

Coverage rewards alignment with phase-wise priors:
\[
\text{Cov}(p_m)=\frac{1}{m}\sum_{i=1}^m P_{\phi_i}(t_{\phi_i}\mid x).
\]

Stealth rewards techniques with lower detection exposure:
\[
\text{Stealth}(p_m)=1-\frac{1}{m}\sum_{i=1}^m \sigma(d_{t_i}).
\]

Exposure penalties discourage techniques with high detection or strong mitigation coverage:
\[
\begin{aligned}
\text{DetPen}(p_m) &= \frac{1}{m}\sum_{i=1}^{m} D(t_i),\\
\text{MitPen}(p_m) &= \frac{1}{m}\sum_{i=1}^{m} M(t_i).
\end{aligned}
\]

A prior penalty reduces over-reliance on Transformer priors:
\[
\text{Prior}(p_m)
  = \frac{1}{m}\sum_{i=1}^{m} 
    \left( 1 - P_{\phi_i}(t_{\phi_i}\mid x) \right).
\]

These components jointly define supervision targets for the Policy--Value Network and serve as the ranking function for final kill-chain hypotheses.

\subsection{Markov Decision Process for Symbolic Transitions}

To capture structured relationships among ATT\&CK phases, we model symbolic transitions using an MDP
\[
\mathcal{M}=(\mathcal{S},\mathcal{A},P,R).
\]
A state $s_i=(\phi_i,t_{\phi_i})$ represents the selected technique for phase $\phi_i$, and the action space from $s_i$ consists of all techniques available in the next phase. Since ATT\&CK does not explicitly encode transition probabilities, we employ a similarity-based kernel:

\[
P(t'\mid t)=
\frac{\exp\left(\alpha\,\text{sim}(t,t')\right)}
     {\sum_{u}\exp\left(\alpha\,\text{sim}(t,u)\right)},
\]

where $\text{sim}$ is cosine similarity over ATT\&CK description embeddings. This induces a soft graph structure where semantically related techniques receive higher transition probability. Algorithm~\ref{alg:mdp} outlines the construction of $P$ and the computation of rollout values $V_{\text{MDP}}$, which approximate long-horizon plausibility and provide value targets for training.

\begin{algorithm}[t]
\DontPrintSemicolon
\KwIn{Technique embeddings $\{v_t\}$, similarity function $\text{sim}$, reward $R(\cdot)$}
\KwOut{Transition kernel $P$, rollout values $V_{\text{MDP}}$}

\ForEach{pair $(t,t')$}{
Compute $s(t,t')=\text{sim}(t,t')$.\;
Compute transition probability:
\[
P(t'|t)=\frac{\exp(\alpha s(t,t'))}
            {\sum_u \exp(\alpha s(t,u))}.
\]
}

\ForEach{state $s_i$}{
Rollout value:
\[
V_{\text{MDP}}(s_i)
=
\mathbb{E}
\left[\sum_{k}\gamma^k R(t_k\!\to\! t_{k+1})\right].
\]
}

\caption{MDP Construction and Rollout}
\label{alg:mdp}
\end{algorithm}

\begin{figure*}[htbp]
    \centering
    \includegraphics[width=\linewidth]{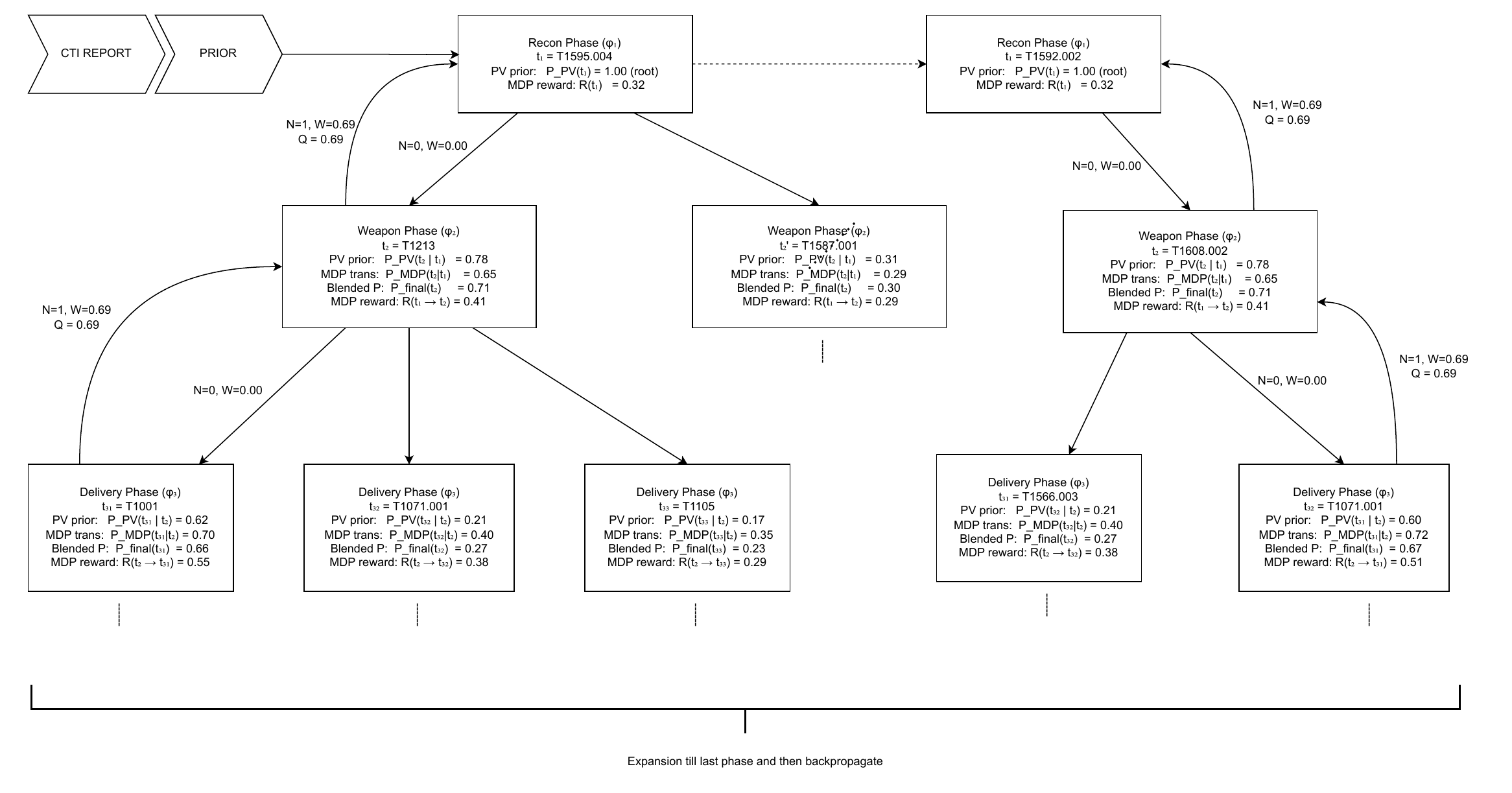}
    \caption{Illustrative MCTS expansion and backtracking trace for the FIN6 report. 
    The diagram shows the partial search trajectory from Recon to Delivery [Iteration 1], 
    highlighting priors and rewards influence node selection.}
    \label{fig:fin6_graph}
\end{figure*}

\subsection{Architectural Rationale and Modeling Principles}

The selection of specific neural components in this framework addresses two fundamental disconnects between natural language narratives and the MITRE ATT\&CK ontology: permutation invariance and conditional modulation.

First, standard sequence models (RNNs or vanilla Transformers) induce an order bias where the position of an input affects its representation. However, the set of candidate techniques available within a specific ATT\&CK phase is inherently unordered; a technique $t_a$ is not ``before'' or ``after'' technique $t_b$ in the vector space. By employing Set Transformers, we enforce permutation invariance, ensuring that the Policy–Value Network aggregates information from the candidate space based solely on semantic utility rather than arbitrary indexing.

Second, the relationship between a CTI narrative and a specific technique is not static but highly context-dependent. A mention of ``Powershell'' in a report could imply Execution, Defense Evasion, or Lateral Movement depending entirely on surrounding semantic cues. The FiLM (Feature-wise Linear Modulation) mechanism is critical here because it allows the global context vector $c$ to affine-transform the feature space of the technique embeddings. Mathematically, this allows the narrative to warp the technique manifold, stretching the distance between irrelevant techniques and compressing the distance between contextually valid ones before the MCTS search begins. This ensures that the search is guided not just by static semantic similarity, but by a dynamic, context-conditioned probability surface.


\subsection{Policy--Value Network}

The Policy--Value Network integrates contextual semantics with symbolic reasoning. Its inputs are the contextual embedding $\tilde{c}$ and the enriched candidate embeddings $\tilde{v}_i$, which undergo two forms of cross-attention:

\[
H_i = \text{Attn}(\tilde{v}_i,\tilde{c},\tilde{c}),\qquad
G   = \text{Attn}(\tilde{c},\{\tilde{v}_i\},\{\tilde{v}_i\}).
\]

The Set Transformer layer then produces permutation-invariant representations
\[
U_i=\text{SetTrans}(H_i),
\]
which encode relationships among candidate techniques independent of ordering. This property is essential because the ATT\&CK technique set at each phase is inherently unordered.

FiLM modulation injects contextual bias:
\[
Z_i=\gamma\odot U_i+\beta,\qquad (\gamma,\beta)=W_f G.
\]

Policy logits are computed as $\ell_i=w_p^{\top}Z_i$, producing a probability distribution via softmax, while the value head aggregates contextual and candidate signals:

\[
v_{\theta}
=
w_v^{\top}
\left(
\text{Pool}(Z_i)\;\|\;G
\right).
\]

The PVN is trained with a joint objective that matches its policy to MCTS visit distributions and its value estimate to rollout-based targets, Algorithm~\ref{alg:pvn} provides the full computation:

\[
\mathcal{L}(\theta)
=
\text{KL}(\pi_{\theta}\,\|\,\pi_{\text{target}})
+
\lambda_v (v_{\theta}-V_{\text{target}})^2.
\]

\begin{algorithm}[t]
\DontPrintSemicolon
\KwIn{Context embedding $c$, candidate embeddings $\{v_i\}$}
\KwOut{Policy $\pi_{\theta}$, value $v_{\theta}$}

Project inputs:
\[
\tilde{c}=W_c c,\qquad
\tilde{v}_i=W_t v_i.
\]

Compute cross-attention:
\[
H_i=\text{Attn}(\tilde{v}_i,\tilde{c},\tilde{c}),\qquad
G=\text{Attn}(\tilde{c},\{\tilde{v}_i\},\{\tilde{v}_i\}).
\]

Set Transformer:
\[
U_i=\text{SetTrans}(H_i).
\]

FiLM modulation:
\[
Z_i=\gamma\odot U_i+\beta.
\]

Policy:
\[
\ell_i=w_p^{\top}Z_i,\qquad
\pi_{\theta}=\text{softmax}(\ell_i).
\]

Value:
\[
h=\text{Pool}(Z_i)\;\|\;G,\qquad
v_{\theta}=w_v^{\top}h.
\]

\caption{Policy--Value Network Computation}
\label{alg:pvn}
\end{algorithm}

\subsection{Monte Carlo Tree Search}

MCTS unifies the semantic priors, symbolic transitions, and value estimates into a principled search procedure. Each node stores $(N,W,Q,P)$ representing visit count, cumulative value, mean value, and prior probability. Selection uses the PUCT rule:

\[
U = Q + c_{\text{puct}}\,P\,\frac{\sqrt{N_{\text{parent}}}}{1+N_{\text{child}}}.
\]

At leaf nodes, the PVN provides a policy $\pi_{\theta}$ and value $v_{\theta}$. Expansion assigns blended priors combining PVN and MDP:

\[
P_{\text{blend}}(t')
\propto
\exp\left(\beta_1 \log \pi_{\theta}(t') + \beta_2 \log P_{\text{MDP}}(t')\right).
\]

Backpropagation updates the statistics, Algorithm~\ref{alg:mcts} summarizes the full process:

\[
N\leftarrow N+1,\quad
W\leftarrow W+v_{\theta},\quad
Q=W/N.
\]

\begin{algorithm}[t]
\DontPrintSemicolon
\KwIn{Root state $s_0$, PV-Net, MDP priors, simulations $S$}
\KwOut{Inferred kill chain}

\For{$k=1$ to $S$}{
Select path using PUCT.\;
Evaluate PV-Net: $(\pi_{\theta},v_{\theta})$.\;
Expand using blended priors $P_{\text{blend}}$.\;
Backpropagate $v_{\theta}$.\;
}

Extract most visited path.\;

\caption{Monte Carlo Tree Search}
\label{alg:mcts}
\end{algorithm}

\subsection{Kill Chain Inference and Behavioral Evaluation}

Given a narrative cyber–threat intelligence report, the objective of the inference module is to reconstruct a seven-phase ATT\&CK kill chain that is semantically consistent with the operational workflow described in the report. The baseline system relies exclusively on a Transformer encoder that maps the narrative to a probability distribution over techniques within each phase. Although this approach captures lexical and local semantic cues, it frequently overestimates techniques that share surface-level terminology with the report while underrepresenting those required for multi-phase operational coherence.

The proposed PV-Network augments the Transformer with a value model and transition dynamics that encode structural constraints between ATT\&CK phases. The model produces, for each phase, a prior distribution conditioned on the preceding technique, and the Monte Carlo Tree Search (MCTS) procedure explores sequences of techniques according to both semantic priors and transition rewards. The system therefore prefers attack paths whose internal dependencies match observed adversary workflows.

To assess the behavioral fidelity of predicted kill chains, we introduce the Nearest-Neighbor Behavioral Alignment (NNBA) score. Let \(H\) denote the set of historical techniques associated with a threat actor and \(P\) the set of predicted techniques for a given intrusion. Each technique \(t\) is represented as a contextual embedding \(\mathbf{e}_t\) generated by ATTACK-BERT. The alignment score is defined as
\[
d_{\mathrm{NN}}(P,H)
= \frac{1}{|P|}
  \sum_{p \in P}
  \min_{h \in H}
  \bigl(1 - \cos(\mathbf{e}_p, \mathbf{e}_h)\bigr).
\]
A lower value indicates that each predicted technique lies close to at least one historically observed technique for the actor, providing an interpretable measure of behavioral plausibility. This metric is used throughout the evaluation and reported alongside qualitative visualizations of the predicted chains.

\begin{table*}[t]
\centering
\caption{ATT\&CK technique predictions across seven phases for three intrusions, with NNBA scores measuring behavioral alignment.}
\label{tab:combined_results}
\resizebox{0.85\textwidth}{!}{
\begin{tabular}{l l l l l}
\toprule
Report & Phase & Transformer Model Path & PV-Network Path & Human Expert Judgment \\
\midrule

\multirow{7}{*}{FIN6 \cite{google_fin6}}
 & Recon      & T1592.002 & T1592.004 & T1087 \\
 & Weapon     & T1570     & T1213     & T1105 \\
 & Delivery   & T1190     & T1001     & T1190 \\
 & Exploit    & T1587.004 & T1587.001 & T1059.001 \\
 & Install    & T1037.003 & T1569.002 & T1569.002 \\
 & C2         & T1552.004 & T1071.001 & T1071.001 \\
 & Objectives & T1558.001 & T1486     & T1486 \\
\midrule

\multirow{7}{*}{APT24 \cite{google_apt24}}
 & Recon      & T1595.003 & T1592.002 & T1595.003 \\
 & Weapon     & T1593.003 & T1608.001 & T1587.001 \\
 & Delivery   & T1001     & T1189     & T1195.002 \\
 & Exploit    & T1608.003 & T1211     & T1203 \\
 & Install    & T1218.007 & T1547.001 & T1547.001 \\
 & C2         & T1546.008 & T1071.001 & T1071.001 \\
 & Objectives & T1558     & T1041     & T1041 \\
\midrule

\multirow{7}{*}{UNC1549 \cite{google_unc1549}}
 & Recon      & T1069     & T1590     & T1598.003 \\
 & Weapon     & T1114.001 & T1608.001 & T1574.001 \\
 & Delivery   & T1190     & T1566.002 & T1199 \\
 & Exploit    & T1588     & T1211     & T1203 \\
 & Install    & T1036.003 & T1574.001 & T1547.001 \\
 & C2         & T1546.008 & T1573     & T1021.004 \\
 & Objectives & T1550.003 & T1213.002 & T1003.006 \\
\midrule

NNBA Score &  & Transformer & PV-Network & Human Expert \\
\midrule
FIN6       &  & 0.1728 & 0.1051 & 0.0133 \\
APT24      &  & 0.2291 & 0.0941 & 0.0902 \\
UNC1549    &  & 0.2253 & 0.0455 & 0.0406 \\
\bottomrule
\end{tabular}}
\end{table*}

\section{Results}

We evaluate the system on three publicly documented intrusions FIN6, APT24, and UNC1549 each of which provides a structured narrative describing attacker objectives, tool choices, and operational sequencing. These narratives serve as semantic references for assessing the fidelity of reconstructed kill chains. Table~\ref{tab:combined_results} lists the predicted techniques for each phase under the Transformer baseline, the proposed PV-Network, and an independent human analyst.

\begin{figure*}[t]
\centering

\subfloat[FIN6 Behavioral Envelope\label{fig:envelope_fin6}]{
    \includegraphics[width=0.31\textwidth]{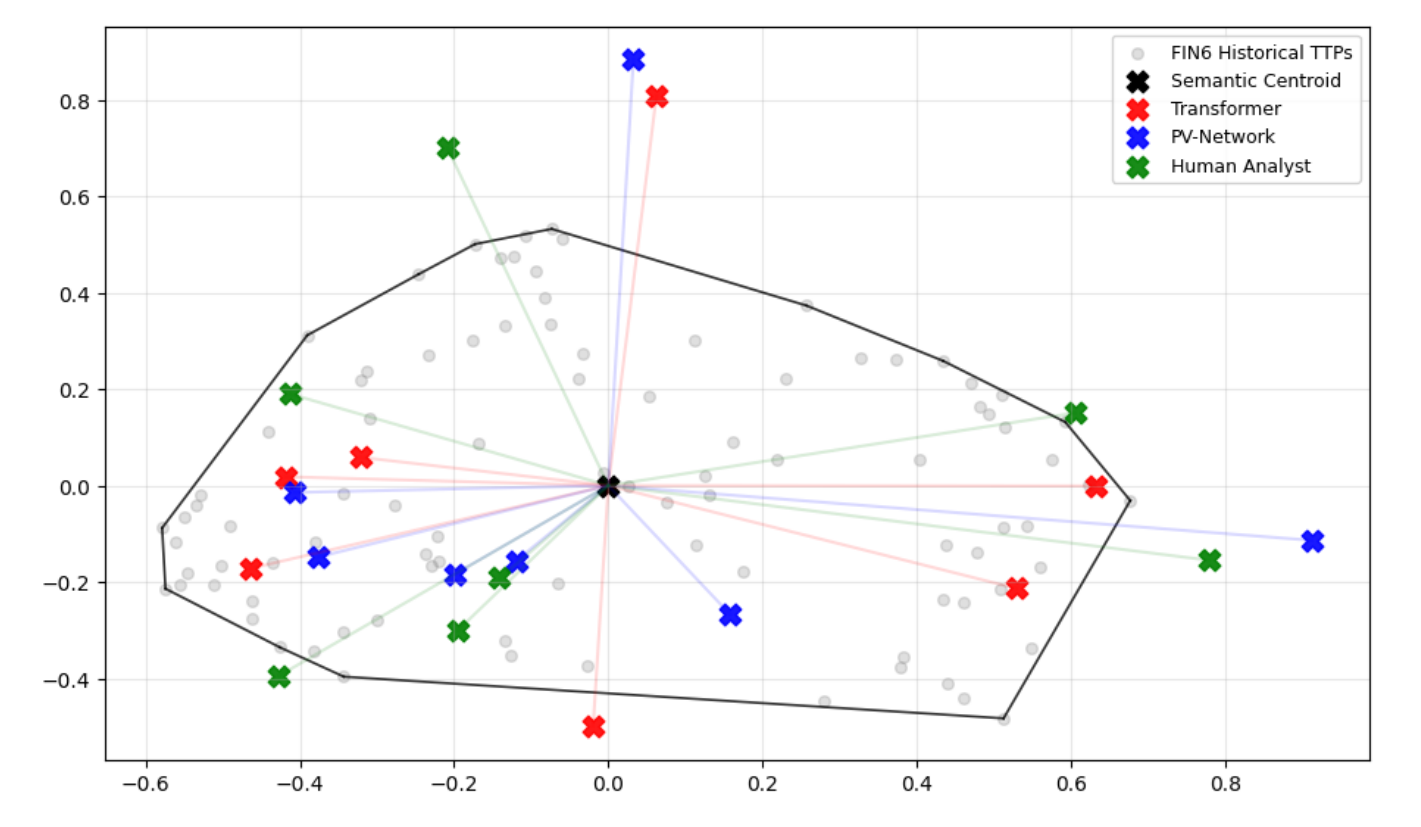}
}
\hfill
\subfloat[APT24 Behavioral Envelope\label{fig:envelope_apt24}]{
    \includegraphics[width=0.31\textwidth]{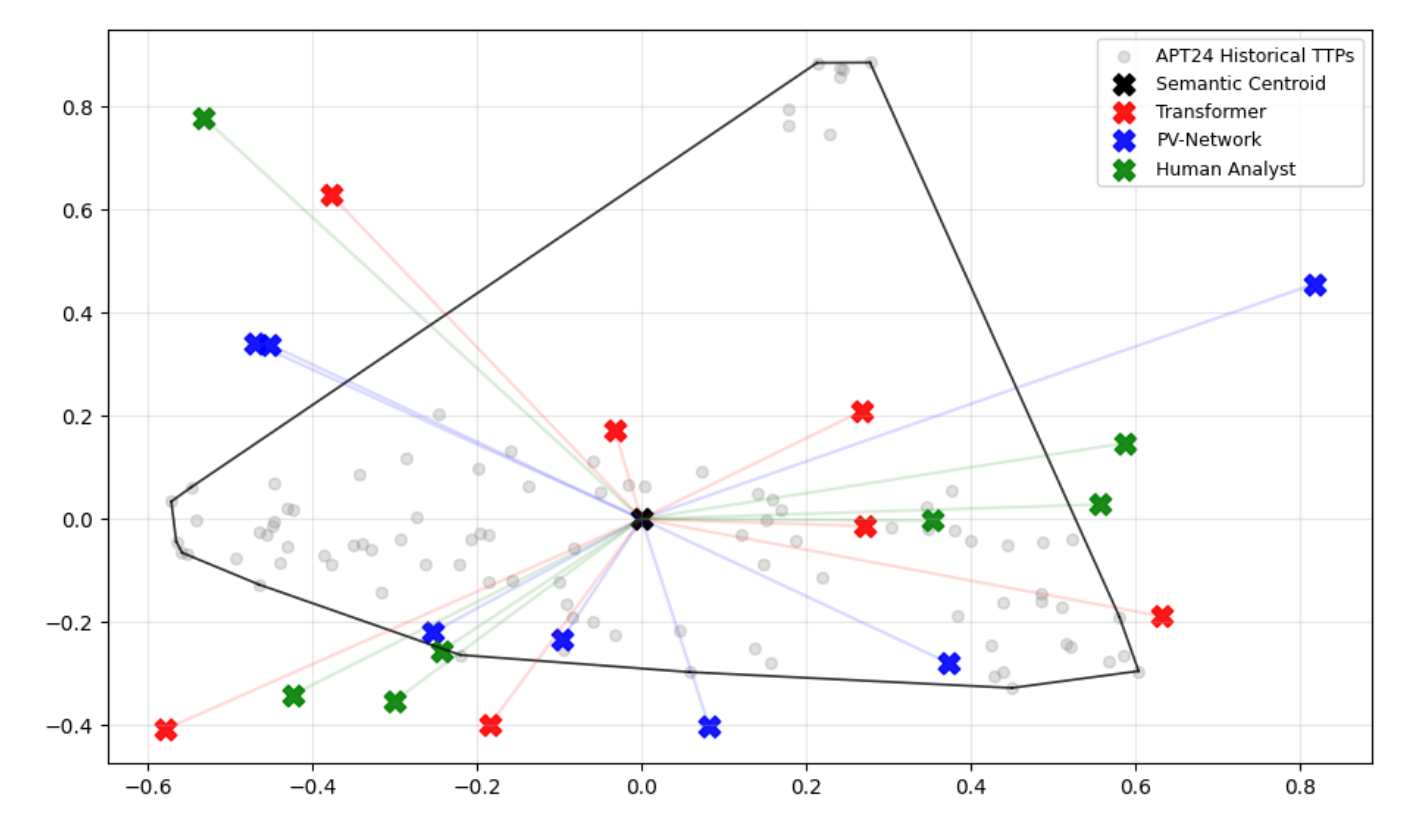}
}
\hfill
\subfloat[UNC1549 Behavioral Envelope\label{fig:envelope_unc1549}]{
    \includegraphics[width=0.31\textwidth]{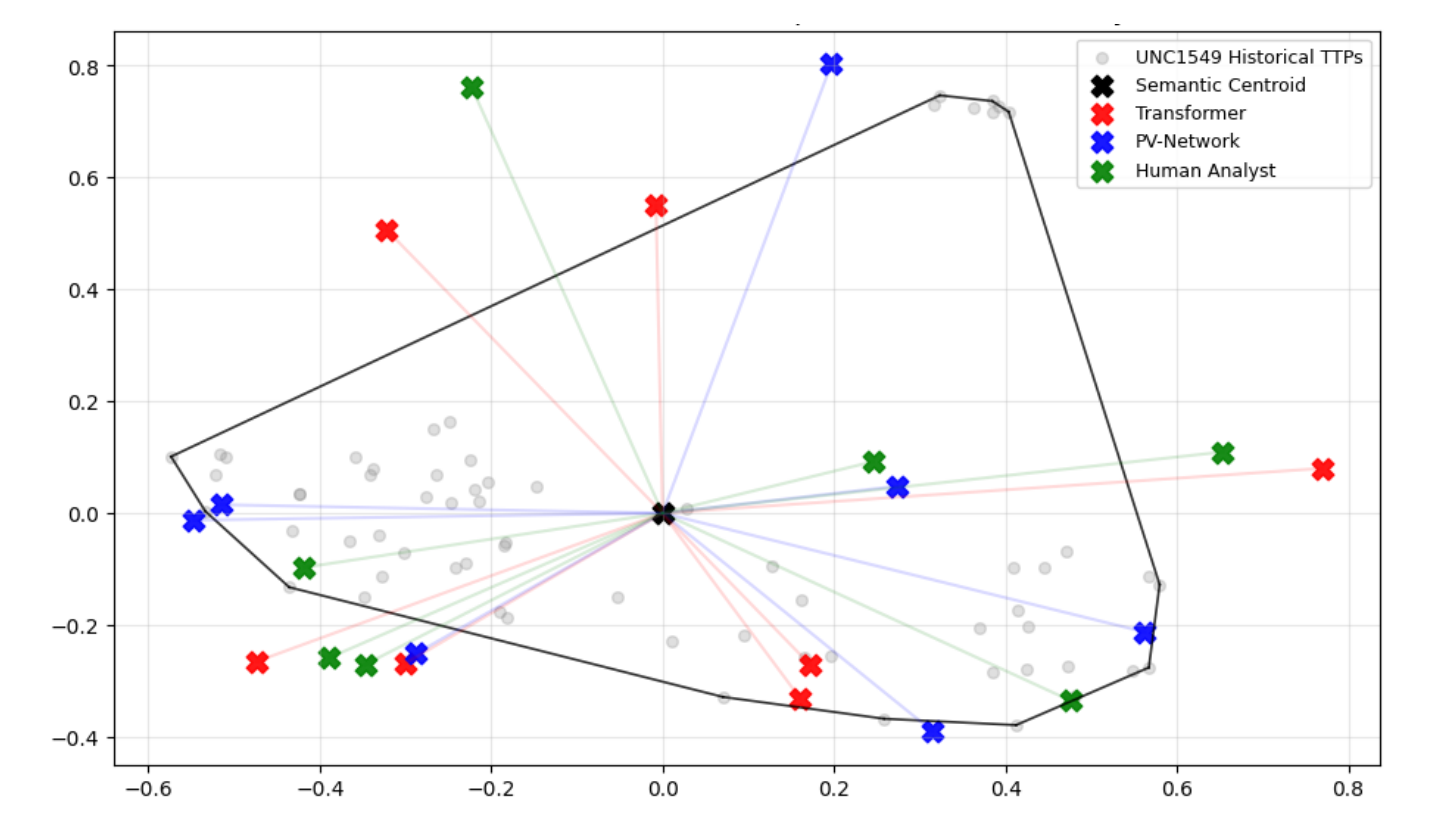}
}

\caption{Behavioral envelope visualizations for threat actors FIN6, APT24, and UNC1549.}
\label{fig:envelopes}
\end{figure*}

To quantify behavioral plausibility, we employ the Nearest-Neighbor Behavioral Alignment (NNBA) score defined in Section~III. This metric compares each predicted technique to the historical ATT\&CK profile of the threat actor and yields an interpretable measure of semantic drift. Across all intrusions, human analysts achieve the lowest alignment scores, reflecting their ability to incorporate broader contextual cues and long-term behavioral priors. The PV-Network consistently outperforms the Transformer baseline, indicating that structural reasoning and reward-guided search substantially improve multi-phase consistency.

Beyond numeric alignment, we analyze the spatial organization of techniques in the embedding space using ATTACK-BERT representations projected through PCA, Fig.\ref{fig:envelopes} depict the convex behavioral envelope for each threat actor constructed from historical techniques, along with the model-predicted techniques positioned relative to the semantic centroid. Techniques that fall inside or near the envelope are more consistent with known operational behaviors, whereas points lying outside suggest semantic deviation.

For FIN6, shown in Fig.\ref{fig:envelope_fin6}, the PV-Network places most predictions near the centroid of historical behavior, particularly for credential harvesting, internal scoping, and covert communication. The Transformer predictions, although occasionally plausible in isolation, form a scattered pattern that departs from the tightly clustered FIN6 envelope. Human predictions remain closest to the centroid and demonstrate the smallest drift.

The APT24 visualization in Fig.\ref{fig:envelope_apt24} reflects the actor's broad operational profile involving staged loaders, web compromises, and phishing infrastructure. The PV-Network captures this structure by placing Weapon and Delivery techniques within the primary behavioral region, while the Transformer tends to produce lexically similar but behaviorally divergent actions. The NNBA scores in Table~\ref{tab:combined_results} confirm this improved alignment.

UNC1549 presents a more diffuse envelope, as seen in Fig.\ref{fig:envelope_unc1549}, stemming from its dual-path operations combining credential theft and execution hijacking. The PV-Network again populates the region consistent with authenticated bypass procedures and staged downloaders, whereas the Transformer frequently selects general exploitation techniques that lie farther from the historical centroid. Human predictions track known UNC1549 practices most closely, reflected in their NNBA score.

Together, the quantitative alignment results and the behavioral envelope visualizations demonstrate that the PV-Network reliably reconstructs multi-phase attack paths that adhere to both semantic cues in the narrative and structural dependencies characteristic of real adversary workflows. While human analysts remain the upper bound of performance, the PV-Network offers a scalable alternative capable of producing behaviorally coherent kill chains across diverse intrusion families.

\section{Discussion}

While the quantitative results demonstrate alignment with historical behaviors, the operational value of the framework lies in its ability to support, rather than replace, human reasoning. A core advantage of the MDP–MCTS architecture over end-to-end black-box classifiers is the transparency of the decision-making process.

Traceability of Inference: Unlike direct classification, where the model outputs a final label without justification, the MCTS procedure generates a search tree that serves as an explanatory artifact Fig.\ref{fig:fin6_graph}. For any predicted technique, an analyst can inspect the back-pointer trace to see which prior techniques (state parents) and which semantic cues (reward signals) led to that specific selection. This allows analysts to distinguish between high-confidence predictions (where the visit count $N$ is concentrated on a single path) and ambiguous scenarios (where $N$ is fractured across diverging hypotheses).

Handling Uncertainty in CTI: CTI reports are often fragmented, describing the impact of an attack (e.g., ``database encrypted'') without explicitly stating the method (e.g., ``SMB propagation''). In such cases, the Policy–Value Network may output a diffuse policy $\pi_\theta$. The MCTS simulation effectively acts as a ``reasoning engine'' that bridges these gaps by favoring transitions that maximize structural coherence ($P_{MDP}$), effectively hypothesizing the most logical bridge between observed states. This allows the system to present analysts with the ``most probable reconstructed path'' while flagging low-confidence links for manual review.

Mitigation of Bias: By decoupling the transition dynamics (learned from ATT\&CK structure) from the semantic extraction (learned from text), the framework reduces the risk of overfitting to specific keywords. A pure language model might hallucinate a technique simply because it frequently co-occurs with a specific malware name in training data. By constraining the search via the MDP transition kernel, our framework filters out hallucinated actions that are linguistically probable but operationally impossible, thereby establishing a higher baseline of trust for automated incident reconstruction.

\section{Conclusion}

This work presented a reasoning driven framework for deriving coherent, ATT\&CK aligned kill chains from unstructured cyber threat intelligence. Unlike static classification approaches, the system integrates contextual semantic encoding, a symbolic MDP for sequential structure, a multi-objective reward model, and an AlphaZero style search procedure guided by a compact Policy-Value Network. This combination enables the model to evaluate partial attack paths not only for linguistic relevance but also for phase-to-phase plausibility and long horizon consistency. Empirical analysis across three real world intrusions shows that the framework produces kill-chain trajectories that more closely reflect operational intent and narrative structure than a Transformer only baseline, and approximate the analytical patterns observed in human expert reasoning.

Future extensions may incorporate partially observable MDP formulations to better accommodate incomplete or noisy intelligence, as well as multi-agent variants to capture attacker defender co-evolution. Integrating threat knowledge graphs offers a path toward richer contextual grounding, while continual learning mechanisms may improve robustness as new TTPs and adversary behaviors emerge. Together, these directions outline a path toward more adaptive and analyst aligned reasoning systems capable of supporting cyber offensive operations under evolving threat conditions.

\bibliographystyle{plain}
\bibliography{bibfile}

\IEEEpubidadjcol 

\end{document}